\documentclass{aa}     
\def\Mpc{$h^{-1}$~Mpc}
\def\hmpc{$h$~Mpc$^{-1}$}

\begin{document}   

   \thesaurus{12 (12.03.3; 12.12.1) }

\title{The supercluster-void network V.}
\subtitle{Alternative evidence for its regularity}

\author {O. Toomet\inst{1}, H. Andernach\inst{2}, J. Einasto\inst{1}, 
M. Einasto\inst{1}, E. Kasak\inst{1}, A. A. Starobinsky\inst{3} \and
E. Tago\inst{1}}

\offprints{J. Einasto }

\institute{ Tartu Observatory, EE-61602 T\~oravere, Estonia
\and
 Depto. de Astronom\'\i a, Univ.\ Guanajuato, 
  Apdo.~Postal 144, C.P.\ 36000, Guanajuato, Mexico
\and
 Landau Institute for Theoretical Physics, Moscow 117 334, Russia}

\date{ Received   1999 / Accepted ...  }

\maketitle

\markboth{Toomet et al.: Regularity of the Universe}{}

\begin{abstract} 

We analyse the distribution of Abell clusters of galaxies to study the
regularity of the supercluster-void network.  We apply a new method
sensitive to the geometry of the location of clusters, and measure the
{\em goodness of regularity} of the network.  We find that the
supercluster-void network resembles a cubical lattice over the whole
space investigated.  The distribution of rich superclusters is not
isotropic: along the main axis of the network it is periodic with a
step of length $\approx 130$~\Mpc, whereas along the diagonal of the
network the period is larger, as expected for a cubical lattice.  This
large-scale inhomogeneity is compatible with recent CMB data.


\end{abstract}

\section{Introduction}

The basic assumption in the standard Friedman-Robertson-Walker
cosmology is that the Universe is homogeneous and isotropic on large
scales (Peebles \cite{p80}).  Clusters of galaxies and galaxy
filaments form superclusters (Abell \cite{abell}); examples are the
Local supercluster (de Vaucouleurs \cite{dv}, Einasto et
al. \cite{ekss84}) and the Perseus-Pisces supercluster (J\~oeveer et
al. \cite{jet78}, Giovanelli \cite{g93}).  Superclusters and voids
between them form a continuous network of high- and low-density
regions which extends over the entire part of the Universe studied in
sufficient detail up to a redshift of $z \approx 0.13$ (Einasto et
al. \cite{me94}, \cite{me97d}, hereafter Paper I). According to the
conventional paradigm the location of superclusters in this network is
random, as density perturbations on all scales are believed to be
randomly distributed (Feldman et al. \cite{fkp}).

There exists growing evidence that the supercluster-void network has
some regularity.  Broadhurst et al.  (\cite{beks}) have measured
redshifts of galaxies in a narrow beam along the direction of the
northern and southern Galactic poles and found that the distribution
is periodic: high-density regions alternate with low-density ones with
a surprisingly constant interval of $128$ \Mpc\ (here $h$ is the
Hubble constant in units of 100~km~s$^{-1}$~Mpc$^{-1}$).  The
three-dimensional distribution of clusters shows clear signs of
regularity (Paper I). One method to characterise this regularity is a
correlation analysis.  Kopylov et al. (\cite{kkf1}, \cite{kkf2}),
Fetisova et al. (\cite{fkl}), Mo et al. (\cite{mo}), and Einasto \&
Gramann (\cite{eg93}) have found evidence for the presence of a
secondary peak of the correlation function of clusters of galaxies at
125~\Mpc.  The secondary peak has been interpreted as the correlation
of clusters in superclusters located on opposite sides of large voids.
Recent studies show that the correlation function oscillates with a
period equal to that of the periodicity of the supercluster-void
network (Einasto et al. \cite{e97b}, hereafter Paper II).  An
oscillation with very low amplitude is seen also in the correlation
function of LCRS galaxies (Tucker et al. \cite{to97}).  An oscillating
correlation function corresponds to a peaked power spectrum (Einasto
et al. \cite{e97c}, hereafter Paper III).  Peacock (\cite{p97}) and
Gazta\~naga \& Baugh (\cite{gb97}) determined the three-dimensional
power spectrum from the projected distribution of APM galaxies; the
Abell cluster power spectrum was derived by Einasto et al.
(\cite{e97a}), and Retzlaff et al.  (\cite{r98}); Tadros et
al. (\cite{tadros}) calculated the power spectrum for APM clusters.
All these recent studies find the peak or turnover of the power
spectrum near a wavelength of $\lambda_0 = 120$~\Mpc\ or wavenumber
$k_0=2\pi/\lambda_0=0.05$~\hmpc.  In addition, in the line-of-sight
correlation function of Ly\,$\alpha$-break galaxies with redshifts
$z\sim 3$ secondary peaks were found at a redshift separation of
$\Delta z \approx 0.22\pm 0.02$ (Broadhurst \& Jaffe \cite{bj99}).
This separation corresponds to the comoving scale $120~(1+z)^{-1}$~Mpc
if a flat $\Lambda$CDM isotropic cosmological model with $\Omega_m =1-
\Omega_{\Lambda}\approx 0.4 \pm 0.1$ is assumed. A low-density model
dominated by a cosmological term fits nicely to many other recent
data, in particular, those from high-$z$ supernovae (Perlmutter et
al. \cite{perl98}).

\begin{figure*}
\vspace*{8.5cm}
\caption{Distribution of clusters in high-density regions in
supergalactic coordinates. Left panel shows clusters in a sheet in
supergalactic $-100 \leq X \leq 200$~\Mpc; Abell and APM clusters
in superclusters with at least 8 or 4 members are plotted with symbols
as indicated.  The supergalactic $Y=0$ plane coincides almost exactly
with the Galactic equatorial plane and marks the Galactic zone of
avoidance. In the right panel only clusters in the southern Galactic
hemisphere are plotted; here the depth is $-350 \leq Y \leq -50$~\Mpc.
}
\includegraphics{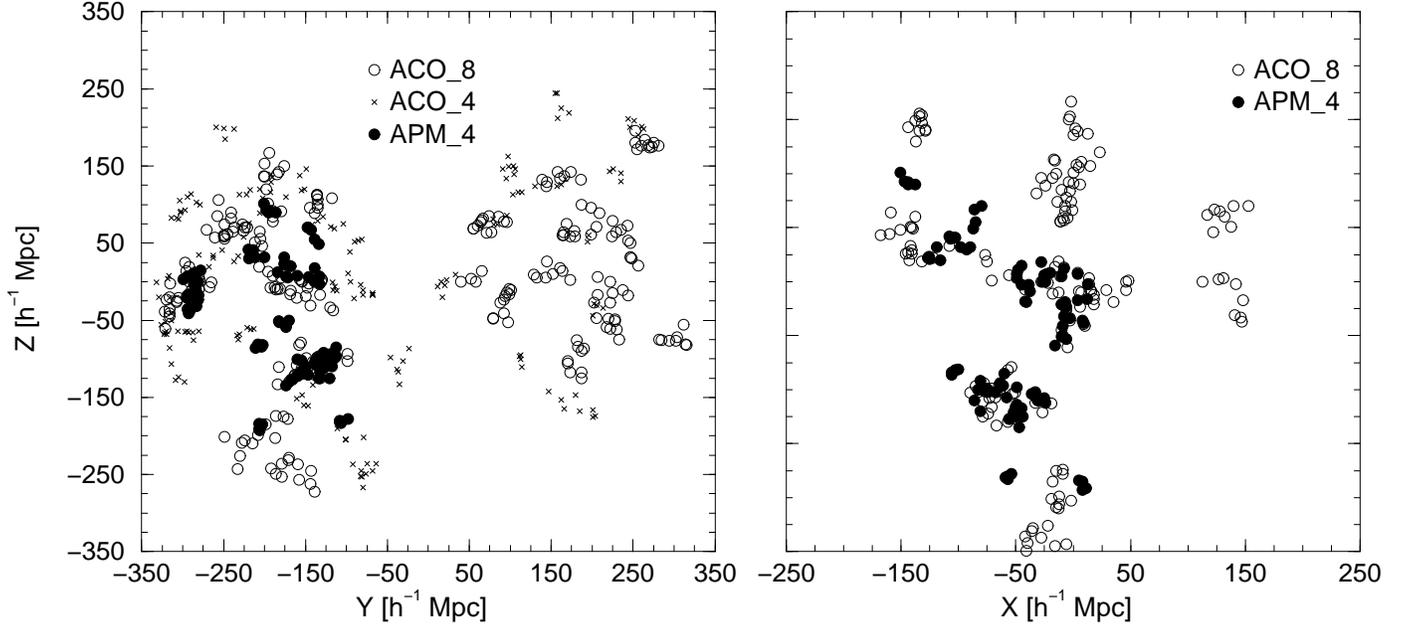}
\label{fig1}
\end{figure*}

It should be noted that no such feature is seen in the power spectra
of some galaxy catalogs, in particular, in the recent IRAS PSC
redshift survey (Sutherland et al. \cite{s99}). However, we suppose
that this may be the result of IRAS galaxies avoiding overdense
regions like rich clusters (see more detailed explanation of this
point by Einasto et al. \cite{e99}).

Quantitative methods used so far to describe the regularity of the
matter distribution are insensitive to the directional and phase
information, and thus characterise the regularity only indirectly.  In
this paper we shall use a new geometric method to investigate the
distribution of clusters of galaxies, sensitive to the regularity and
the isotropy of the distribution.

\section{Observational data}

As in previous papers of this series we use the 1995 version of the
compilation by Andernach \& Tago (\cite{at98}) of all published galaxy
redshifts towards galaxy clusters in the catalogue of rich clusters of
galaxies by Abell (\cite{abell}) and Abell et al. (\cite{aco})
(hereafter Abell clusters).  Individual galaxies were associated to a
given Abell cluster if they lay within a projected distance of $\leq
1.5$~\Mpc\ (1 Abell radius) and within a factor of two of the redshift
estimated from the brightness of the clusters 10-th brightest galaxy,
using the photometric estimate of Peacock \& West (\cite{pw92}).  For
the present analysis we used a sample of all rich clusters (richness
class R$\ge 0$ and excluding clusters from ACO's supplementary list
of S-clusters) in this compilation with redshifts up to $z=0.12$.
The sample contains 1304 clusters, 869 of which have measured
redshifts for at least two galaxies. Distances of clusters without
measured redshifts and of clusters with only 1 galaxy measured have
been estimated on the basis of the apparent magnitude of the 10-th
brightest galaxy of the cluster.

As representatives of high-density regions in the Universe we used
{\em rich\,} superclusters from the list of superclusters presented in
Paper I.  Superclusters were identified using the friend-of-friends
algorithm by Zeldovich et al.  (\cite{zes82}) with a neighbourhood
radius of 24~\Mpc. In this way all clusters of a supercluster have at
least one neighbour at a distance not exceeding the neighborhood
radius.  To illustrate the distribution of clusters in high-density
regions we plot in Fig.~1 only clusters in rich superclusters, while
in the quantitative analysis below we shall use all clusters.  The
sheets plotted are 300~\Mpc\ thick, thus some superclusters overlap in
projection.  The sheet in the left panel of Fig.~1 crosses the
majority of cells of the supercluster-void network present in our
cluster survey; the sheet in the right panel has the full depth of the
sample in the southern Galactic hemisphere.  Fig.~1 shows clearly the
quasi-regular network of superclusters interspersed with voids.  The
three-dimensional distribution of all Abell and APM clusters in rich
superclusters in the whole space within a limiting radius 350~\Mpc\
around us is shown in the home page of Tartu Observatory
(http://www.aai.ee).

For comparison we plot in Fig.~1 also the distribution of APM clusters
of galaxies in rich superclusters. We see that the APM cluster sample
covers a much smaller volume in space which makes it difficult to
investigate the regularity of the distribution of high-density regions
on large scales. The APM cluster sample is defined only in the
southern Galactic hemisphere, and even here the APM sample containing
clusters with measured redshifts covers only three high-density
regions defined by very rich superclusters with at least 8
member-clusters. These are the Sculptor (SC9), Pisces-Cetus (SC10),
and Horologium-Reticulum (SC48) superclusters of the catalogue in
Paper I.  To investigate the regularity of the supercluster-void
network, the sample volume must exceed the period of the network at
least several times.  For this reason we used in the following
analysis only Abell clusters.

\begin{figure} 
\vspace{12.5cm}  
\caption
{The main idea of the method. The sample volume (upper panel) is
divided into smaller trial cubes (dotted lines) and the contents of
all test cubes is stacked together (lower panels). If the side-length
of the trial cubes is equal to the real period, there will be a clear
density enhancement (lower left panel) in the stacked distribution.
Otherwise the stacked distribution is almost random (lower right
panel). The size of the elementary sub-cell within the trial cube is
shown in the upper left corner of the stacked cell. }
\includegraphics{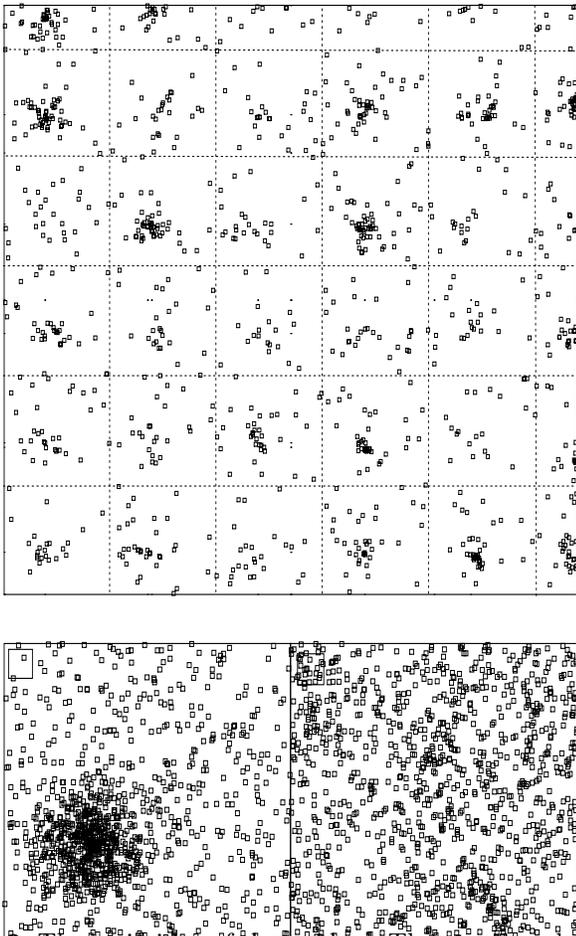}
\label{fig2}
\end{figure}

\section {Method}

To investigate the regularity of the distribution of clusters of
galaxies we shall use a novel geometric method. The idea of the method
is taken from the periodicity analysis of time series of variable
stars. To derive the period of a variable star the time series of
brightness measurements is folded using a certain trial period, i.e.
individual measurements are stacked to yield a combined light curve
for a single period.  If the trial period is wrong then the
distribution of data-points in the combined curve is almost random.
In contrast, if the trial period is correct and the star under study
is a periodic variable, then one has a clear mean light curve of the
star.

This method has been generalised to the 3-dimensional case by Toomet
(\cite{ott}).  The whole space under study is divided into trial cells
of side length $d$.  Then all objects in the individual cubical cells
are stacked to a single combined cell, preserving their phases in the
original trial cells. We then vary the side length of the trial cube
to search for the periodicity of the cluster distribution.  The method
can be used for any form of tightly packed regular trial cells. We use
cubical trial cells which have the simplest possible form, and because
this form gives a satisfactory explanation of the distribution of
clusters.

\begin{figure*} 
\vspace{7cm}
\caption 
{Curves of goodness of regularity for quasiregular mock
samples. Samples have 300 randomly located clusters and 42 (left
panel) or 100 (right panel) clusters in quasiregularly located
superclusters (for details of construction of mock samples see Paper
III).  The step of the regular cubical grid is $r_0=130$~\Mpc; the
samples extend over the whole cubical volume.  In the left panel we
show the case in which the axes of the trial cubes are oriented at
$\theta=0^{\circ}$ with respect to the major axis of the regular
network; the $2\sigma$ error corridor is marked with thin dashed
lines. In the right panel the solid, dashed and dot-dashed lines are
for trial cubes oriented at $\theta=0^{\circ}$, $22.5^{\circ}$ and
$45^{\circ}$ with respect to the major axis of the regular network,
respectively.  }
\includegraphics{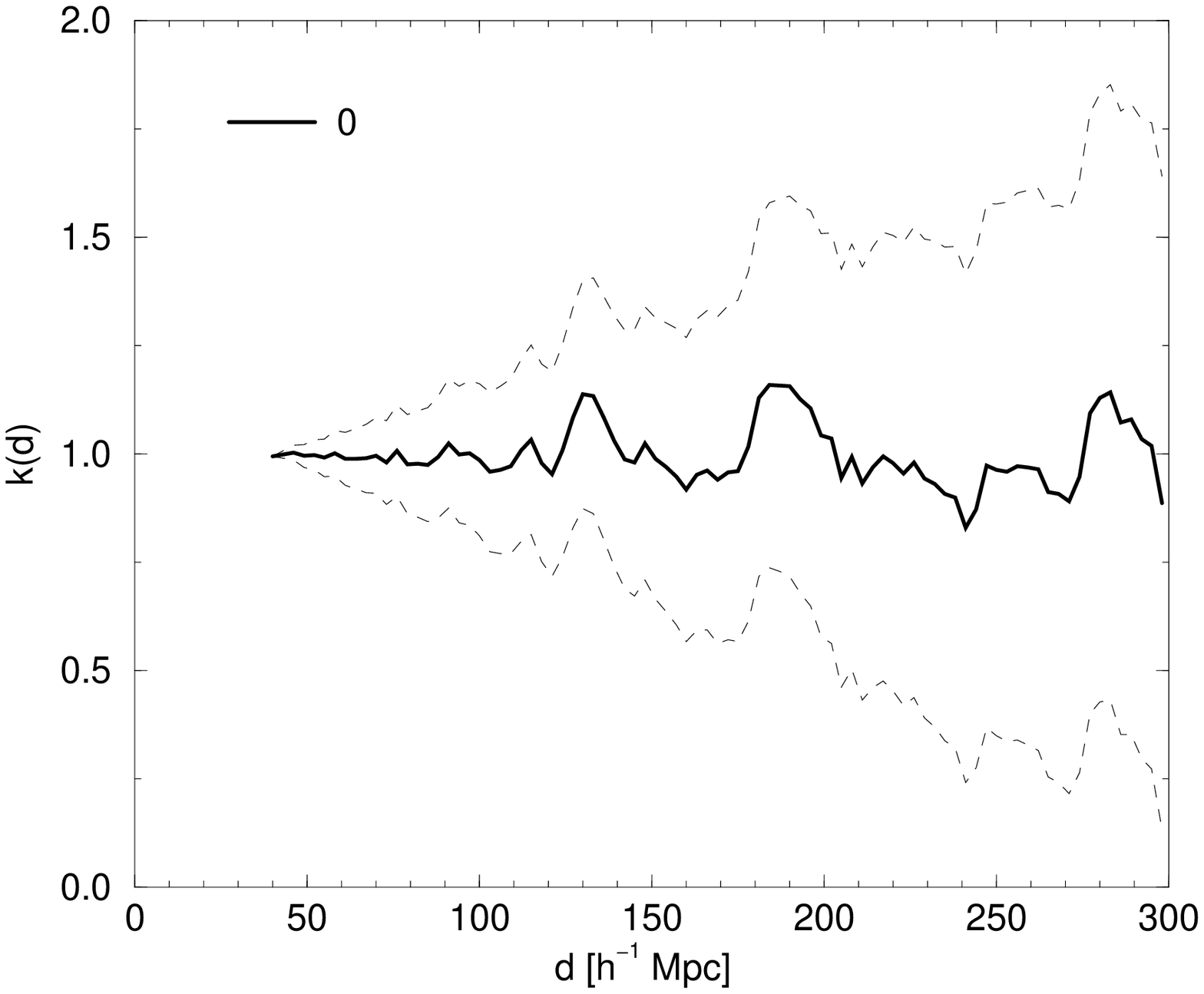}
\includegraphics{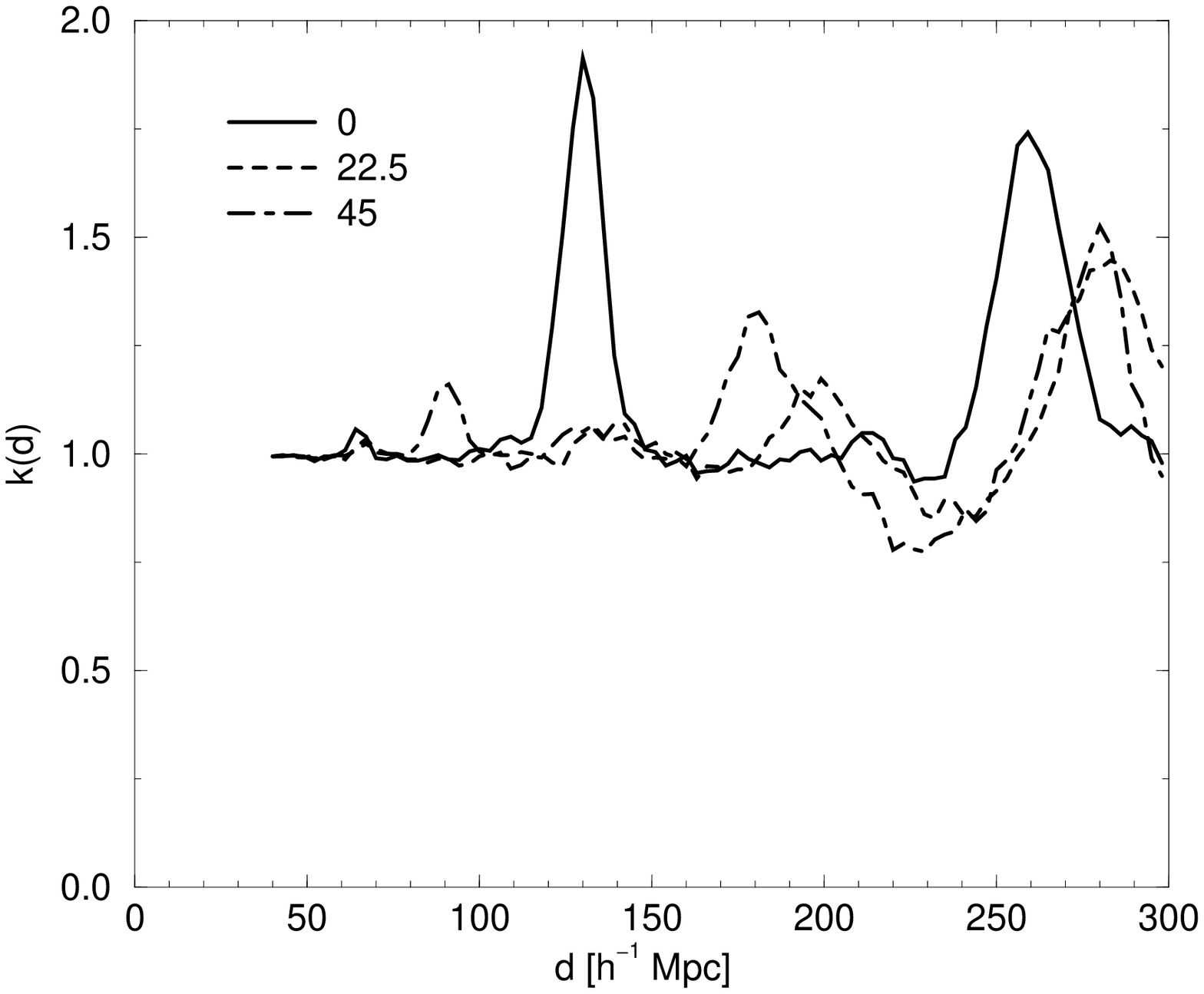}
\label{fig3}
\end{figure*} 

\begin{figure*}
\vspace{7cm}
\caption{Curves of goodness of regularity for observed cluster
samples.  In the left panel we show the case $\theta=0^{\circ}$
i.e.\ trial cubes oriented along supergalactic coordinates; thin
dashed lines show the $2\sigma$ error corridor.  In the right panel we
show the resulting curves of goodness of regularity for trial cubes
oriented at $\theta=0^{\circ}$, $22^{\circ}$ and $45^{\circ}$ with
respect to supergalactic coordinates.}
\includegraphics{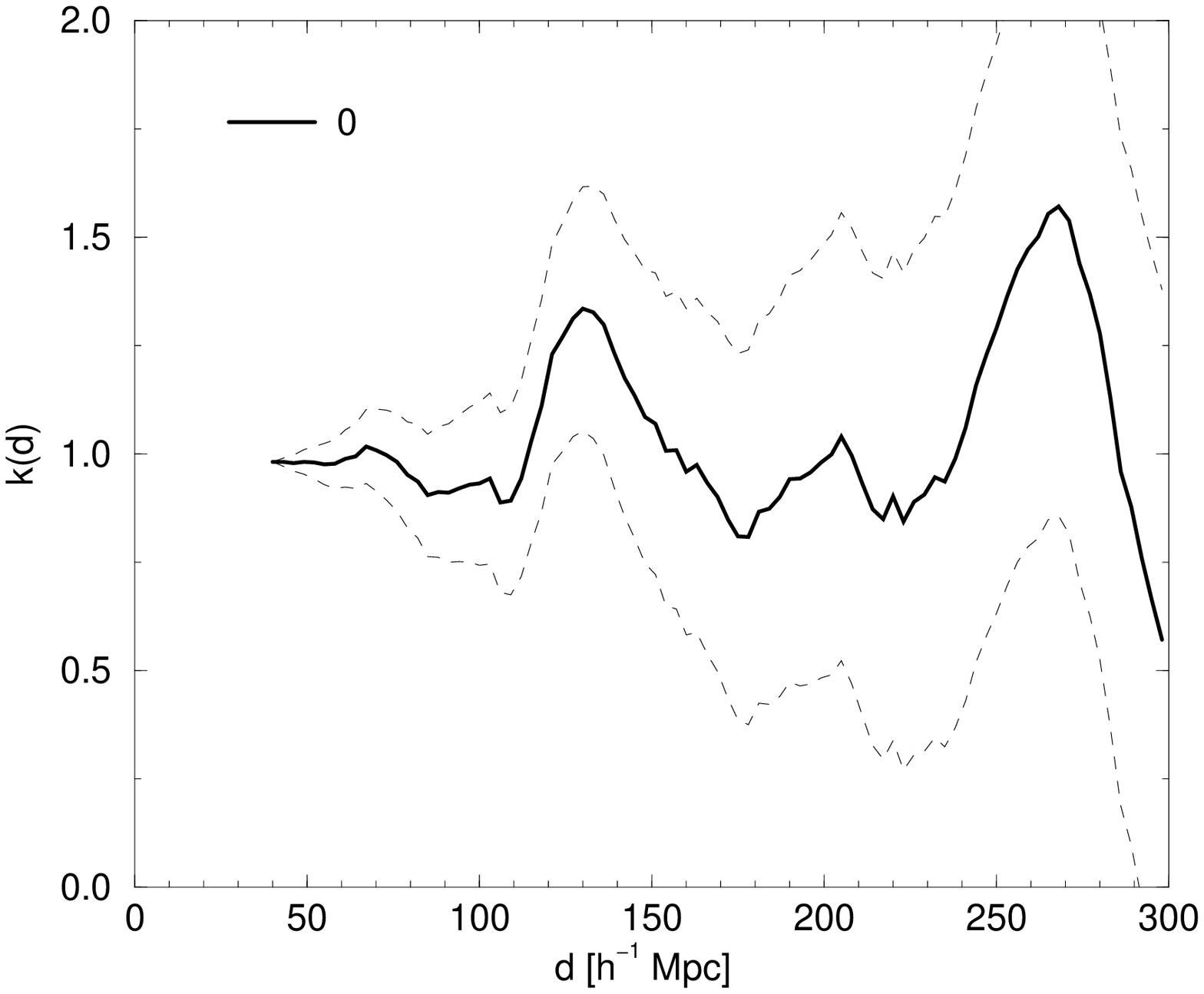}
\includegraphics{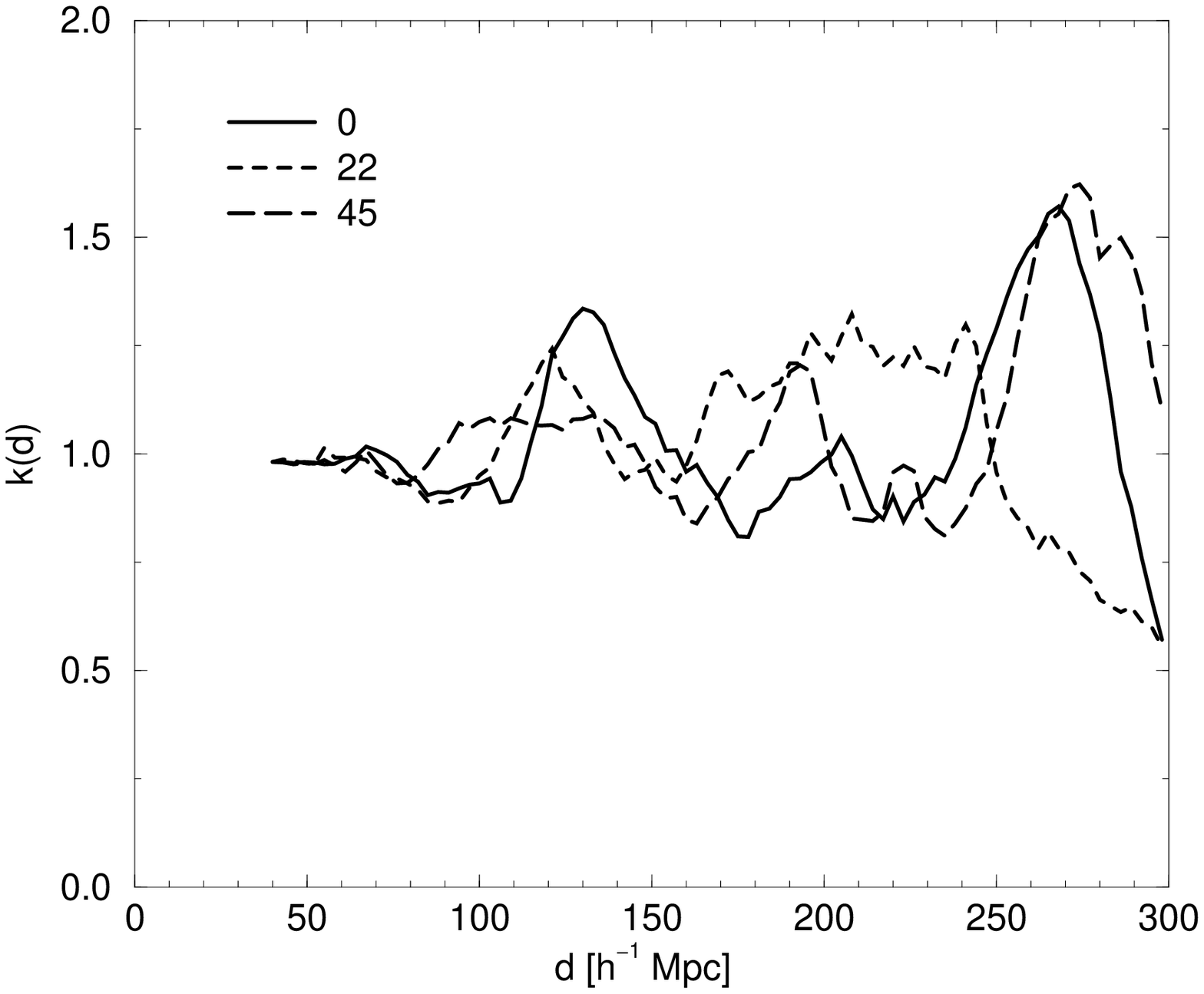}
\label{fig4}
\end{figure*}

The next step is to calculate the {\em goodness of regularity} for
this side length $d$ of the trial cell.  To do this we divide the
combined stacked cell into small elementary sub-cells and count, for a
particular data point (i.e.\ cluster of galaxies) within this
sub-cell, the number of {\em other} data points which come from an
original trial cell {\em different} from the one in which the given
data point is located.
The {\em goodness of regularity} is calculated as follows:
\begin{equation}
   k(d) = {2\over{ N(N + 1)}}{V(d)\over{V_\varepsilon}}\sum_{i=1}^N
   \sum_{j=i+1}^N v(\vec f_i, \vec f_j).
\label{ott}
\end{equation}
Here $N$ is the total number of data points, i.e.\ individual Abell
clusters; $V\equiv d^3$ is the volume of the trial cell of side length
$d$; $\vec f_i$ is the phase-vector of the cluster $i$ in the combined
cell (a spatial remainder-vector, the components of which are
remainders, which arise if the original coordinates are divided by
$d$); $V_\varepsilon$ is the elementary volume inside of the combined
cell. We use an elementary cube (sub-cell) of side length
$2d_\varepsilon$.  The counter $v( \vec f_i,\vec f_j)$ is defined as
follows: $v_{ij} = 1$, if the phase vector of the cluster $j$ lies in
the elementary volume $V_\varepsilon$ (the centre of which coincides
with $\vec f_i$), and if clusters $i$ and $j$ are located farther away
from each other than $d_\varepsilon$.  Otherwise we set $v_{ij} = 0$.
Thus $v_{ij}$ is unity only if the phases of clusters are close enough
in the combined cube, but the clusters themselves are not members of
the same supercluster (they lie in different trial cubes of the
original space).

The sensitivity of the method and the rms error of the variable $k(d)$
have been determined using mock samples of randomly and
quasi-regularly located points.  We use mock samples with two
populations of clusters, in one population all clusters are randomly
distributed, in the second population they are located in
superclusters which form a regular rectangular network with a step
130~\Mpc.  Supercluster centre positions have random shifts $\pm
20$~\Mpc\ around corners of the regular network.  Mock samples are
described in more detail in Paper III.  Here we are interested in the
cosmic error due to variance of $k(d)$ in different realizations of
the sample.  This relative error can be expressed as $\sigma_k \approx
b (d - 2 d_{\varepsilon})/N$, where $N$ is the number of clusters in
the sample, and $b\approx 0.5$ is a constant determined from the
analysis of mock samples.  The dependence of the error on $d$ comes
from the fact, that, if $d = 2d_{\varepsilon}$, then trial cells have
the same size as elementary sub-cells. Thus all clusters of the trial
cell are located in the elementary sub-cell and there is no variance
of $k(d)$ between different realizations; for larger trial cell size
$d$ the error increases.  The presence of a periodicity in the sample
can be determined with confidence if the amplitude of the goodness
near the maximum deviates from the Poisson value $k=1$ by more than
$2\sigma$.

Results of the analysis of mock samples are illustrated in Figure~3.
We see in the right panel that the goodness curve has well defined
maxima at integer multiples of $r_0=130$.  In the mock sample shown in
the left panel the fraction of clusters located in regular network is
smaller and the amplitude of the maximum lies within the $2\sigma$
error corridor. This example shows that the period of a regular
network of points can be determined if the population of regularly
distributed points is sufficiently large (at least 10~\% of the total
number of points) and if the total number of particles is at least
several hundred.

Obviously, the method is sensitive to the direction of the axes of the
trial cubes.  If clusters form a quasi-rectangular cellular network,
and the search cube is oriented along the main axis of the network,
then the period is found to be equal to the side length of the cell.
If the search cube is oriented at some non-zero angle $\theta$ with
respect to the major axis of the network, then the presence of a
periodicity and the period depend on the angle.  If the angle
$\theta\approx 45^{\circ}$, then the period is equal to the length of
the diagonal of the cell.  If the angle differs considerably from
$\theta=0^{\circ}$ and $45^{\circ}$, the periodicity is weak or
absent. This property is also illustrated in Figures 3 and 4 for mock
and real cluster samples, respectively.

Results of the period analysis of our cluster sample are shown in
Figure~4.  The axes of the search cubes were oriented at various
angles with the coordinate axes of the supergalactic coordinate
system.  As seen from these Figures, the period really depends on the
orientation.  The smallest value of the period, $P=130$ \Mpc, occurs
when the cubes are oriented along supergalactic coordinates.  If the
trial cube is oriented at $\theta=45^{\circ}$ with respect to
supergalactic coordinates, then the period is $P=190$ \Mpc, as
expected for a rectangular network of cells. In both cases there exist
a second maximum of the curve of goodness of regularity at $d=250~
-280$~\Mpc\ which is also well pronounced.  If the trial cube is
oriented at $22^{\circ}$, then the periodicity is much weaker, the
first maximum at $d=120$~\Mpc\ has a lower amplitude, and the second
is absent.  In principle there are more possibilities, for instance
when the trial cells are oriented along the diagonal of the cubical
lattice. However, in most additional cases the periodicity is weaker.
Probably this can be due to the fact the the regularity is not very
strong and there exist deviations from it. Much larger samples are
needed to find limits of the regularity.

\section{Discussion and conclusions }

The periodicity analysis confirms the analysis made in Papers I and II
using other methods. The cluster sample has a considerable fraction of
clusters located in rich superclusters which form a rectangular
lattice with a period of about $120 - 130$~\Mpc.  Our analysis shows
that the supercluster-void network is oriented approximately along
supergalactic coordinates.  This confirms earlier results on the
presence of a high concentration of clusters and superclusters towards
both the Supergalactic Plane (Tully et al. \cite{tu3}), and towards
the Dominant Supercluster Plane, which are at right angles with
respect to each other (Paper~I).  Supergalactic $Y$ axis is very close
to the direction of Galactic poles, thus it is natural to expect a
well-defined periodicity along Galactic poles as indeed observed by
Broadhurst et al. (\cite{beks}).  The rectangular character of the
distribution of rich clusters was also noticed by Tully et al.
(\cite{tu3}).  Recently Battaner (\cite{b98}) has found that many
known superclusters can be identified with the vertices of a 
octahedron network of the superstructure.

One principal result of our study is the direction dependence of the
periodicity.  This conclusion was not possible with the power spectrum
or correlation function approach since they are not sensitive to
directional information.  A clear periodicity is observed only along
supergalactic coordinates.  This is in good agreement with results by
Broadhurst et al. (\cite{beks}) where a periodicity was observed in
the direction of Galactic poles, i.e.\ almost exactly along the
supergalactic $Y$ axis (supergalactic $X$ and $Z$ axes lie close to
the galactic plane where clusters are invisible).  In other directions
the regularity is less pronounced (Guzzo et al. \cite{gcnl}, Willmer
et al. \cite{wksk}, Ettori et al. \cite{egt}).  As noted already by
Bahcall (\cite{b91}), the nearest peaks of the Broadhurst et al. survey
coincide in position and redshift with nearby rich superclusters.
Thin deep slices, such as slices of the LCRS and the Century Survey,
also show  a weak periodicity signal (Landy et al. \cite{ls96},
Geller et al.  \cite{g97}).  The scale length found in these studies
is of the same order as derived in the present paper.

The regularity found in this paper concerns high-density regions
marked by clusters in rich superclusters. When we compare the
distribution of galaxies in various regions of the supercluster-void
network we see considerable individual differences.  On small scales
the distribution of galaxies and clusters is fractal, as shown by many
studies (see Wu et al. \cite{wu99}); and is well reproduced by models
with randomly distributed density perturbations (see Feldman et
al. \cite{fkp}).  On very large scales covering the whole volume of
the Abell cluster sample we see in the distribution of clusters in
extremely rich superclusters with $N> 30$ members no regularity.  We
come to the conclusion that a clear regularity is observed only near
the wavelength $\lambda_0 \approx 120$~\Mpc\ of the peak of the power
spectrum.  This result can be expressed as follows: density
perturbations on very small ($k \gg k_0$) and very large wavelengths
($k \ll k_0$) are uncorrelated and have random phases; however, near
the wavelength $\lambda_0 = 2\pi/k_0$ they are correlated and have
similar phases.  The importance of the use of phase information was
stressed by Szalay (\cite{szalay}).

The regularity of the large-scale structure of the Universe has been
studied also using the distribution of centres of superclusters
(Kalinkov et al. \cite{kal}, Kerscher \cite{ker}).  Kalinkov et
al. searched for the presence of a high-order clustering of
superclusters using the correlation analysis, and found that
superclusters are not clustered.  Our analysis in Papers I and II and
presented here suggests that the regularity is completely different:
it consists of the presence of the supercluster-void network, not the
clustering of superclusters (like galaxies concentrate to clusters).
The supercluster-void regularity is expressed by the distribution of
clusters themselves, not supercluster centres.  Kerscher used a
combination of the nearest neighbour distribution and the void
probability function to measure the regularity of the structure.  This
method works well for small separations between superclusters ($r \leq
60$~\Mpc), and yields results in good agreement with Paper I.
However, the method does not characterise the supercluster-void
network on larger scales.

It should be emphasized that this inhomogeneity in the Universe does
not contradict the observed degree of isotropy of the cosmic microwave
background (CMB) radiation. As follows from general expressions for
{\em rms} values of multipoles $C_l$ of CMB angular temperature
anisotropies (see, e.g., Starobinsky \cite{st88}), any additional
localized excess of power in the Fourier spectrum of density
perturbations introduced at a scale $k_0\ll 1/R_H$ affects only
multipoles $l<l_0=k_0R_H$ where $R_H$ is the present horizon radius.
This is valid for both Gaussian and non-Gaussian additional
perturbations (assuming only that there is no correlation between
perturbations with $k\approx k_0$ and perturbations from the other,
``regular'' part of the spectrum).  Moreover, multipoles with
$l=(0.7-0.9)l_0$ are mainly amplified. For the pure CDM cosmological
model with $\Omega_m=1$, $R_H=2c/H_0$ and $l_0=300$ mainly the first
acoustic peak is enhanced (Atrio-Barandela et al. \cite{a97};
Eisenstein et al. \cite{Eis98}; Broadhurst \& Jaffe \cite{bj99}).
However, in the $\Lambda$CDM model (which is strongly supported by
numerous recent observational data) this enhancement shifts to larger
values of $l$. In particular, if $\Omega_m = 1 - \Omega_{\Lambda} =
0.3$, then $R_H=3.305c/H_0$, and $l_0\approx 500$. For this model, the
effect discussed in this paper mainly results in an increase of $C_l$
in the valley between the first and second acoustic peaks of the CMB
temperature anisotropy spectrum.  In particular, the amplitude
$(\Delta T/T)_l = \sqrt{C_l~l(l+1)/2\pi}$ for $l=400$ will be $\approx
2\times 10^{-5}$, in agreement with recent CAT2 data (Baker et
al. \cite{b99}), as well as with previous CAT1 results (Scott et
al. \cite{s96}). A more detailed treatment of resulting CMB
anisotropies in the presence of a cosmological constant will be
presented in a separate paper.

The main conclusion we can draw from our study is that the Universe is
not homogeneous and fully isotropic on scales $\approx 120$~\Mpc.
High-density regions in the Universe form a quasi-rectangular lattice.
The distribution of high-density regions depends on the direction:
along the main axis (coinciding with the supergalactic $Y$ axis) of the
cellular system the regularity is well pronounced and has a period of
$120 - 130$~\Mpc, while in other directions the distribution is less 
regular.

\begin{acknowledgements} 

We thank Jaan Pelt for discussion and suggestions.  The present study
was supported by Estonian Science Foundation grant 2625.  A.S. was
partially supported by the grant of the Russian Foundation for Basic
Research No. 99-02-16224. This paper was finished during his stay at
the Institute of Theoretical Physics, ETH, Zurich.

\end{acknowledgements}

\end{document}